\documentclass[aps,prl,amssymb,superscriptaddress,10pt,tightenlines,twocolumn,showpacs]{revtex4-1}

\usepackage[english]{babel}
\usepackage{graphicx}
\usepackage{dcolumn}
\usepackage{bm}
\usepackage{color}
\usepackage{subfigure}
\usepackage{amsfonts}
\usepackage{amsmath}
\usepackage{amssymb}
\usepackage{hyperref}

\def\bra#1{\langle #1 |}
\def\ket#1{| #1 \rangle}

\def\e{\mathrm{e}}

\def\e{\mathrm{e}}
\def\ii{\mathrm{i}}
\def\dd{\mathrm{d}}

\newcommand{\tr}{\mathop{\text{Tr}}\nolimits}

\begin{document}

\title{Long-lived entanglement of two multilevel atoms in a waveguide}

\author{Paolo Facchi}
\affiliation{Dipartimento di Fisica and MECENAS, Universit\`{a} di Bari, I-70126 Bari, Italy}
\affiliation{INFN, Sezione di Bari, I-70126 Bari, Italy}

\author{Saverio Pascazio}
\affiliation{Dipartimento di Fisica and MECENAS, Universit\`{a} di Bari, I-70126 Bari, Italy}
\affiliation{Istituto Nazionale di Ottica (INO-CNR), I-50125 Firenze, Italy}
\affiliation{INFN, Sezione di Bari, I-70126 Bari, Italy}

\author{Francesco V. Pepe}
\affiliation{Museo Storico della Fisica e Centro Studi e Ricerche ``Enrico Fermi'', I-00184 Roma, Italy}
\affiliation{INFN, Sezione di Bari, I-70126 Bari, Italy}

\author{Kazuya Yuasa}
\affiliation{Department of Physics, Waseda University, Tokyo 169-8555, Japan}

\date{\today}

\begin{abstract}
We study the presence of nontrivial bound states of two multilevel quantum emitters and the photons propagating in a linear waveguide. We characterize the conditions for the existence of such states and determine their general properties, focusing in particular on the entanglement between the two emitters, that increases with the number of excitations. We discuss the relevance of the results for entanglement preservation and generation by spontaneous relaxation processes.
\end{abstract}


\maketitle

\textit{Introduction.}---The physics of quantum systems confined in one-dimensional (1D) geometries has recently attracted a lot of attention~\cite{Giamarchi,Kuramoto}, and is motivating interesting theoretical and experimental research. The behavior of an excited atom coupled to a field is among the peculiarities of such systems: although decay by spontaneous emission occurs in free (three-dimensional) space, boundary conditions and artificial dimensional reduction alter the picture, enhancing or inhibiting (and sometimes hindering) decay. These effects have been extensively studied and observed in cavity-QED settings~\cite{QED1,QED2,QED3,decay1,decay2,decay3,decay4,decay5,decay6,cavity1,cavity2,cavity3}, where the spectrum of the electromagnetic field is discrete. It is much less trivial that similar phenomena occur in effectively 1D unbound systems, in which the field spectrum is continuous and photons are free to propagate in 1D space. Dimensional reduction can be implemented in a range of experimental platforms, that include cold atoms in tightly focused fields~\cite{focused1,focused2,focused3}, photonic crystals~\cite{kimble1,kimble2,onedim1,onedim2}, optical fibers~\cite{onedim3,onedim4}, quantum dots in photonic nanowires~\cite{semiinfinite1,semiinfinite2}, and superconducting qubits in integrated circuit waveguides~\cite{onedim5,onedim6,mirror1,atomrefl1}. Theoretical studies focused on the interplay between the spectral features of the field and the structure of the emitters~\cite{cirac1,cirac2,cirac3,CCCR,JXGR,CalRabl,3levels1,3levels2}.

The vacuum of quasi-1D fields and their coupling with quantum emitters (real or artificial atoms) can be engineered by properly adjusting the distance between the emitters and a perfect mirror at one end of the system~\cite{mirror1,mirror2,mirror3}. However, the interplay between absorption, stimulated and spontaneous emission provides a quantum emitter with mirror-like properties~\cite{atomrefl1,atomrefl2,atomrefl3}. Hence, a pair of emitters can confine the field in the region between them, yielding nontrivial bound states above the threshold for photon propagation, that can be exploited for their robust entanglement features~\cite{twoatoms1,twoatoms2,twoatoms3,twoatoms4,twoatoms5,waveguide_pra}.

\begin{figure}
\centering
\includegraphics[width=0.35\textwidth]{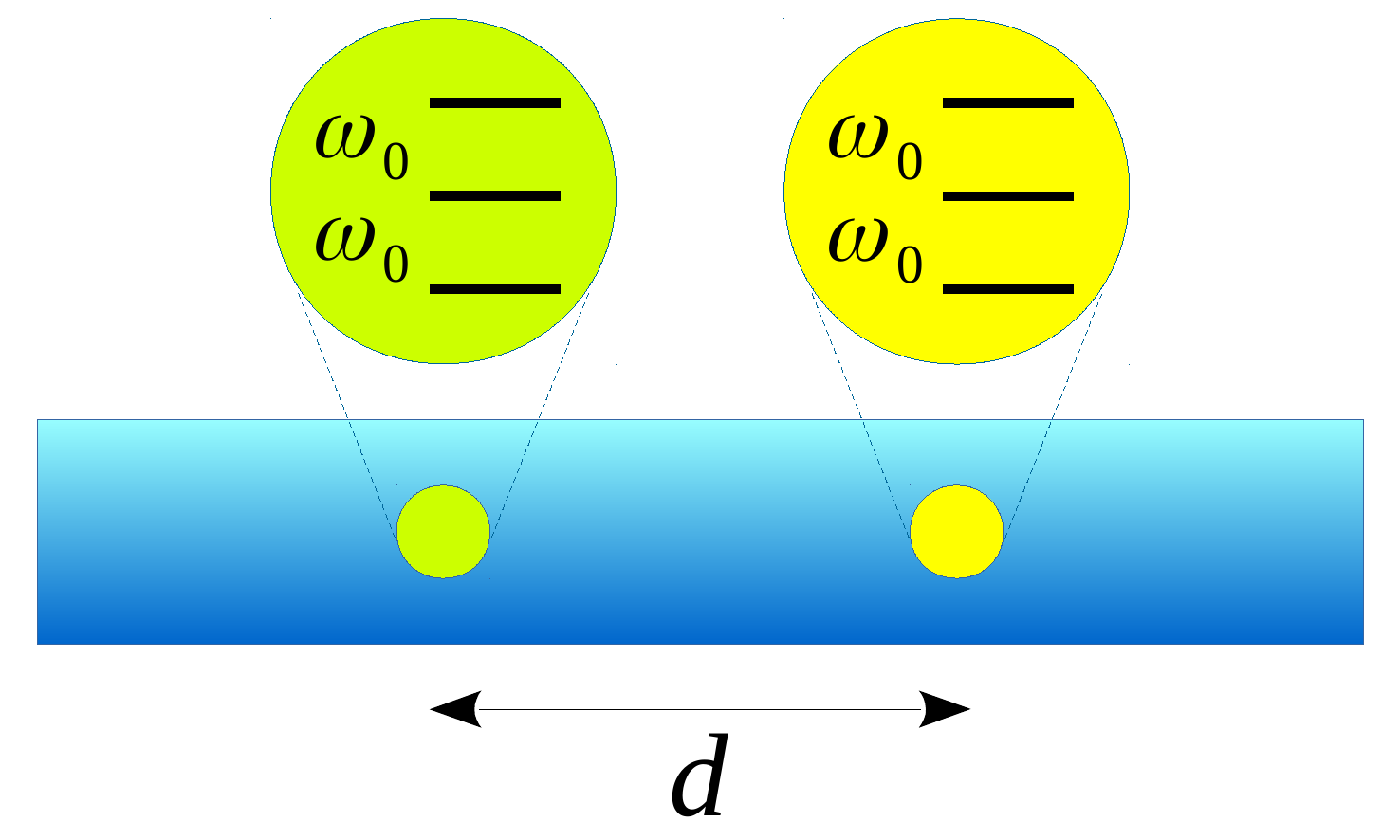}
\caption{A pair of distinguishable $N$-level atoms with equally spaced levels, placed at a distance $d$ in a 1D cavity. $N=3$ in the figure. The case $N =+\infty$ corresponds to a pair of harmonic oscillators.}
\label{fig:qutrits}
\end{figure}

The objective of this Letter is to study the possible stable configurations of two multilevel atoms placed in a 1D cavity. See Fig.~\ref{fig:qutrits}.
In analogy with the 2-level case, each atom behaves both as an emitter and a mirror, confining the photon field and giving rise to a bound state endowed with highly nontrivial entanglement between the two atoms and between the atoms and the field. The effects to be discussed are non-perturbative and enable entanglement generation by relaxation. We will adopt a method of resolution that applies to a generic multilevel emitter with equally spaced energy levels. The technique consists in solving the problem for full-fledged harmonic oscillators, endowed with infinite number of levels. Due to the rotating-wave form of the Hamiltonian and the ensuing conservation law, the evolution in a given sector will involve only a finite number of atomic excitations.

\textit{Model.}---We consider a pair of emitters consisting of distinguishable harmonic oscillators $A$ and $B$, with the same characteristic frequency $\omega_0$, placed in a linear waveguide at a distance $d$. We assume that the photons coupled to the oscillators belong to a single nondegenerate transverse mode of the waveguide, with dispersion relation $\omega(k)$. Hence, at the zeroth order in the coupling constant, the oscillator frequency must be larger than the low-energy cutoff \begin{equation}
M:=\min_k \omega(k) \geq 0
\end{equation}
to enable propagation along the guide, and smaller than the other mode cutoffs to justify the assumption of coupling to a single mode. These conditions can be typically realized in a linear rectangular waveguide, where the dispersion relation of the lowest-energy mode reads $\omega(k)=(k^2+M^2)^{1/2}$, with $M$ inversely proportional to the longer side of the guide cross-section. However, we will keep the discussion as general as possible. In the dipolar and rotating-wave approximations, the Hamiltonian reads
\begin{eqnarray}
\label{hamiltonian}
H
& = & \omega_0 (b^{\dagger}_A b_A + b^{\dagger}_B b_B) + \int \dd k\, \omega(k) b^{\dagger}(k) b(k) \nonumber \\ 
& & {} + \int \dd k\, g(k)\, \Bigl[ (b^{\dagger}_A + b^{\dagger}_B \e^{\ii kd}) b(k) + \mathrm{H.c.} \Bigr] ,
\end{eqnarray} 
where $b(k)$ and $b^{\dagger}(k)$ are the photon field operators in longitudinal momentum space, satisfying the canonical commutation relation $[b(k),b^{\dagger}(k')]=\delta(k-k')$, while $b_J$ and $b^{\dagger}_J$ ($J=A,B$) are the canonical harmonic oscillator operators, satisfying $[b_J,b^{\dagger}_K]=\delta_{JK}$. The real coupling function $g(k)$ naturally decouples at high frequencies,  such that 
\begin{equation}
\label{eq:low-pass}
\int \frac{g(k)^2}{1 + \omega(k)} \dd k <+\infty
\end{equation}
(see e.g.~\cite{waveguide_pra} for photon waveguides). 

The excited (number) states are created by acting on the vacuum $\ket{0}=\ket{0_A,0_B}\otimes\ket{\mathrm{vac}}$ with the creation operators
$b_A^\dagger$, $b_B^\dagger$, and $b(k)^{\dagger}$.
The Hamiltonian~(\ref{hamiltonian}) commutes with the total number of excitations
\begin{equation}\label{number}
\mathcal{N} = \mathcal{N}_{\mathrm{at}} + \mathcal{N}_{\mathrm{field}} = b^{\dagger}_A b_A + b^{\dagger}_B b_B + \int \dd k\, b^{\dagger}(k) b(k) 
\end{equation}
and does not mix different sectors, belonging to different values of $\mathcal{N}$. Due to this conservation law and the robustness of our approximations~\cite{NPSY}, our analysis applies equally well to a pair of harmonic oscillators and to a pair of $N$-level atoms (with equally spaced levels) in a waveguide. See Fig.~\ref{fig:qutrits}.

\textit{Bound states.}---The $\mathcal{N}=0$ sector contains only the vacuum $\ket{0}$. In the $\mathcal{N}=1$ sector, for discrete values of the interatomic distance, one finds the existence of an atom-photon bound state, which has no counterpart in more than one dimension. We shall prove that the presence of this bound state is fundamental in determining the properties of highly-excited sectors.

Since the Hamiltonian is quadratic in the field operators, it can be diagonalized by a proper linear combination of the bosonic field operators. In particular, consider the generic combination
\begin{equation}\label{S}
b_\phi = \phi_A b_A + \phi_B b_B + \int \dd k \, \phi(k) b(k),
\end{equation}  
with
\begin{equation}
|\phi_A|^2 + |\phi_B|^2 + \int \dd k\, |\phi(k)|^2 = 1.
\end{equation}
$b_\phi$ is a bosonic annihilation operator and satisfies the canonical commutation relation $[b_\phi,b_\phi^{\dagger}]=1$. Suppose it satisfies the equation
\begin{equation}\label{ladder}
[b_\phi,H]= Eb_\phi, 
\end{equation}
for some real $E$. Then it is immediate to see that $b_\phi^{\dagger}$ creates \textit{bound} eigenstates of $H$ from the vacuum. Moreover, applying it $N$ times to the vacuum will create eigenstates of $H$, belonging to the sector $\mathcal{N}=N$, with eigenvalues $NE$.

By~(\ref{S}), equation~(\ref{ladder}) is equivalent to
\begin{eqnarray}
E \phi_A & = & \omega_0 \phi_A + \int \dd k\, g(k) \phi(k), \label{eigenvalueA} \\ 
E \phi_B & = & \omega_0 \phi_B + \int \dd k\, g(k) \phi(k) \e^{\ii kd}, \label{eigenvalueB} \\ 
\phi(k) & = &  g(k) \frac{\phi_A + \phi_B \e^{-\ii kd}}{E - \omega(k)}. \label{eigenvalueF}
\end{eqnarray}
If the system admits a solution with $E<M\leq \omega(k)$, then the amplitude $\phi(k)$ is square integrable and the state is also normalizable, provided  condition~(\ref{eq:low-pass}) holds.
These solutions correspond to bound states below the threshold for photon propagation: therefore, they occur even in the case of a single excited emitter coupled to the waveguide. 

The solutions above threshold ($E>M$) are more interesting and nontrivial, since in such conditions a single excited emitter would spontaneously decay by photon emission. In fact, this happens in most situations also for a pair of  emitters. Normalizability of $\phi(k)$ requires that the poles $k_i$ in~\eqref{eigenvalueF}, solutions to $E=\omega(k)$, are compensated by zeros in the numerator: thus,  $g(k)$ or $\phi_A+\phi_B \e^{-\ii kd}$ must vanish at all poles. Let us exclude the former possibility and focus on the latter: since 
\begin{equation}
\phi_A+\phi_B \e^{-\ii k_i d} = 0,
\end{equation} 
for all $i$, in order to obtain a nontrivial solution, the poles must be constrained by the conditions $(k_i - k_j)d=2\pi n$, with $n$ an integer. Plugging these results into Eqs.~(\ref{eigenvalueA})--(\ref{eigenvalueB}) must yield a real solution $E$, which does not depend on the choice of the pole $k_i$. This sets a strong limitation to the possibility of bound states above threshold. A solution can in principle be found when both $\omega(k)$ and $g^2(k)$ are symmetric in $k$, and $\omega$ is increasing in $|k|$, in which case $k_1=-k_2=\bar{k}$ and the energy $E=\omega(\bar{k})$ of the bound state must satisfy the coupled equations
\begin{eqnarray}
\omega(\bar{k}) & = & \omega_0 + \int \dd k \, g^2(k) \frac{1- \e^{\ii(\bar{k}-k)d}}{\omega(\bar{k})-\omega(k)}, \\
\bar{k} d & = & n \pi, \quad \text{with}\ n\in\mathbb{N}. \label{resonance}
\end{eqnarray}
The conditions on the existence of a bound state in this case imply $\phi_A=(-1)^{n+1}\phi_B$, so that the probability $p_{\mathrm{at}}=2|\phi_A|^2$ associated to the emitter component of $b_\phi$ reads
\begin{equation}\label{prob}
p_{\mathrm{at}} = \left( 1 + \int \dd k \, g^2(k) \frac{1-\cos[(\bar{k}-k)d]}{[\omega(\bar{k})-\omega(k)]^2} \right)^{-1} .
\end{equation}
This quantity differs from unity by an order $\lambda^2 \equiv \|g\|^2$, but can become small for $\omega(\bar{k})$ close to the threshold $M$~\cite{waveguide_pra}.

\textit{Entanglement.}---We can now study the properties of the bound states, that occur, at the lowest order in the emitter-photon coupling $\lambda$, whenever the distance between $A$ and $B$ satisfies $d\simeq n\pi/|\omega^{-1}(\omega_0)|$, with $n$ a positive integer. The bound state can be expanded as
\begin{align}
\label{boundstate}
\ket{N} &= \frac{(b_\phi^{\dagger})^N}{\sqrt{N!}} \ket{0} \nonumber \\ & = \sum_{m=0}^N \begin{pmatrix} N \\ m \end{pmatrix}^{\frac{1}{2}} p_{\mathrm{at}}^{\frac{m}{2}} (1-p_{\mathrm{at}})^{\frac{N-m}{2}} \ket{\psi^{(m)}_{AB}} \otimes \ket{\phi^{(N-m)}} ,
\end{align}
where we have introduced the normalized states of the emitters
\begin{align}
\label{psiAB}
\ket{\psi^{(m)}_{AB}} & := \frac{1}{\sqrt{p_{\mathrm{at}}^m m!}} \left( \phi_A^* b_A^{\dagger} + \phi_B^* b_B^{\dagger} \right)^m \ket{0_A,0_B} \nonumber \\
& = 2^{-\frac{m}{2}} \sum_{\ell=0}^m \begin{pmatrix} m \\ \ell \end{pmatrix}^{\frac{1}{2}} (-1)^{\ell(n+1)} \ket{\ell_A, (m-\ell)_B}
\end{align}
and of the photon field
\begin{equation}
\ket{\phi^{(m)}} := \frac{1}{\sqrt{m!}} \left[ (1-p_{\mathrm{at}})^{-\frac{1}{2}} \int \dd k\, \phi^*(k) b^{\dagger}(k) \right]^m \ket{\mathrm{vac}} .
\end{equation}
State $\ket{N}$, that is characterized by a fixed number of total excitations, has components in the emitter excitation sectors ranging from $\mathcal{N}_{\mathrm{at}}=0$, correlated with the presence of $N$ photons, to $\mathcal{N}_{\mathrm{at}}=N$, with no photon in the guide. When the coupling $\lambda$ is sufficiently small and $p_{\mathrm{at}} \simeq 1$, the bound state is dominated by the contribution from $\mathcal{N}_{\mathrm{at}}=N$, namely
\begin{equation}
\ket{N} = p_{\mathrm{at}}^{\frac{N}{2}} \ket{\psi^{(N)}_{AB}} \otimes \ket{\mathrm{vac}} + O(N \lambda^2).
\label{psiNAB}
\end{equation}
Notice that the state $\ket{\psi^{(N-1)}_{AB}}$, that is dominant in $\ket{N-1}$, appears in the $O(N\lambda^2)$ term of Eq.~(\ref{psiNAB}). 

Two comments are in order. First, the state belongs to the $\mathcal{N}=N$ sector and never leaves it, under the action of the Hamiltonian~(\ref{hamiltonian}). Thus, although the preceding analysis has been done for two harmonic oscillators, it is still valid for two $N$-level systems in the waveguide. This point will be formalized in the final part of the article. Second, the reduced density matrix of the two emitters 
\begin{align}
\rho^{(N)}_{AB} & = \tr_{\mathrm{field}} \ket{N}\bra{N} \nonumber \\ & = \sum_{m=0}^N \begin{pmatrix} N \\ m \end{pmatrix} p_{\mathrm{at}}^{m} (1-p_{\mathrm{at}})^{N-m} \ket{\psi^{(m)}_{AB}} \bra{\psi^{(m)}_{AB}}
\end{align}
is a mixture of entangled states, dominated by the term $m=N$, whose entanglement increases with the number of excitations. Before quantitatively clarifying the last point, let us comment on the consequences and possible applications of entanglement in a bound state.

The interesting properties related to entanglement persistence stem from the decomposition of the Hamiltonian~(\ref{hamiltonian}) into a ``stable'' and a ``decaying'' part, as
\begin{equation}
H = E b_\phi^{\dagger} b_\phi + H_c,
\end{equation}
where $H_c$, that commutes with $b_\phi$, accounts for the continuous spectrum and the spontaneous decay of excited states orthogonal to $\ket{N}$. Here we have assumed that there is only one resonant bound state in the spectrum, which is valid in the small coupling regime. In the one-excitation sector (namely two 2-level atoms), it has been demonstrated~\cite{waveguide_pra} that the presence of the bound state enhances the decay rate of the orthogonal states towards the configuration in which the emitters are both in their ground states and a photon propagates in the waveguide. 

If the emitters are placed at infinite distance from one another, an initial state close to $\ket{\psi^{(N)}_{AB}}\otimes\ket{\mathrm{vac}}$ with $N>0$ would rapidly decay to an orthogonal state with a smaller number of emitter excitations. By contrast, in a waveguide, with $d$ and $\omega_0$ close to the resonance conditions, such a state will be left almost invariant by the Hamiltonian evolution, with a slight dressing due to the imperfect superposition of the initial and the bound state (i.e.~$p_{\mathrm{at}}<1$). Therefore, on timescales smaller than the waveguide losses, entanglement is preserved \emph{without} imposing constraints or external control. 

Another interesting application is related to the decay of the unstable component of an \textit{arbitrary} initial state. Indeed, an initial state $\rho_{\mathrm{in}} \otimes \ket{\mathrm{vac}} \bra{\mathrm{vac}}$, with $\rho_{\mathrm{in}}$ in the $\mathcal{N}_{\mathrm{at}}=N$ sector,  will relax towards the $\ket{\psi^{(N)}_{AB}}$ component of the bound state with probability
\begin{equation}
p_{\mathrm{in}} = p_{\mathrm{at}}^{2N} \bra{\psi^{(N)}_{AB}} \rho_{\mathrm{in}} \ket{\psi^{(N)}_{AB}} ,
\end{equation}
with the other terms of the asymptotic density matrix decaying to the sectors with $\mathcal{N}_{\mathrm{at}}<N$ or coupling different sectors. This strategy is similar to the Hamiltonian generation of entanglement, in which an initial (factorized) state is let to coherently evolve until it reaches an entangled state. Such a procedure can be applied in the case of the bound states well below the threshold for photon propagation. However, the Hamiltonian nature of the evolution in the $AB$ Hilbert space yields oscillations, which implies that the evolution must be stopped at a proper time to obtain the desired state. This drawback is absent in entanglement generation by relaxation, in which the final state is approached asymptotically. This technique does not require energy pumping into the system, since a constant entanglement is reached after an initial transient~\cite{waveguide_pra}.

Let us go back to the discussion of the entanglement properties of $\ket{\psi_{AB}^{(N)}}$. The emitter state in the sector $\mathcal{N}_{\mathrm{at}}=1$ (two 2-level atoms) has been extensively studied, and is particularly interesting since it corresponds to one of the two Bell states, according to the sign of $(-1)^n$ [see Eqs.~(\ref{resonance})--(\ref{psiAB})]. It is thus maximally entangled in the $\mathcal{N}_{\mathrm{at}}=1$ subspace. 
The bound state $\ket{N=2}$, relative to the pair of 3-level atoms in Fig.~\ref{fig:qutrits}, reads
\begin{align}
\label{boundstate2}
\ket{2} ={} & \frac{p_{\mathrm{at}}}{\sqrt{6}}\,\Bigl( \ket{0_A,2_B} - 2 \ket{1_A,1_B} + \ket{2_A,0_B} \Bigr) \otimes \ket{\mathrm{vac}} \nonumber \\
 & {}+ \sqrt{2 p_{\mathrm{at}}(1-p_{\mathrm{at}})}\,\Bigl( \ket{0_A,1_B} - \ket{1_A,0_B} \Bigr) \otimes \ket{\phi^{(1)}} \nonumber \\
 & {}+ (1-p_{\mathrm{at}}) \ket{0_A,0_B} \otimes \ket{\phi^{(2)}}.
\end{align}
By projecting onto a suitable photonic state, one can select the desired (long-lived) atomic entangled state $\ket{\psi_{AB}^{(N)}}$, for $N=1$ and 2.
To extend the analysis of entanglement to large-$N$ states, we can use the fact that the reduced one-emitter density matrices obtained from $\ket{\psi_{AB}^{(m)}}$ have a particularly simple binomial form, that leads to
\begin{equation}\label{rhoAN}
\rho_A^{(N)} = \tr_B\rho_{AB}^{(N)} = \sum_{\ell=0}^N C_{\ell}^{(N)} (p_{\mathrm{at}}) \ket{\ell_A} \bra{\ell_A},
\end{equation}
where the coefficients
\begin{equation}
C_{\ell}^{(N)} (p_{\mathrm{at}}) := \sum_{m=0}^N \frac{1}{2^m}  \begin{pmatrix} N \\ m \end{pmatrix}\begin{pmatrix} m \\ \ell \end{pmatrix}p_{\mathrm{at}}^{m} (1-p_{\mathrm{at}})^{N-m} 
\end{equation}
are dominated by the terms $m=N$ at small coupling. These states appear in the analysis of coherently illuminated beam splitters, when the input states are very imbalanced~\cite{NPSY}, and their interesting entanglement features generalize those of NOON states.
Bipartite entanglement can be quantified through any measure based on the eigenvalues of the reduced state $\rho_A^{(N)}$. Let us adopt purity of $A$: 
\begin{align}\label{purity}
\pi_{A}^{(N)} & = \sum_{\ell=0}^N \Bigl(C_{\ell}^{(N)} (p_{\mathrm{at}})\Bigr)^2 = \frac{\Gamma\!\left(N+\frac{1}{2}\right)}{\sqrt{\pi} N!} \,\Bigl( 1 + O[ (N\lambda^2)^2] \Bigr) \nonumber \\
&  
\sim \left[\frac{1}{\sqrt{\pi N}} + O(N^{-3/2}) \right] \Bigl(1 + O[ (N\lambda^2)^2] \Bigr),
\end{align}
as $N\to\infty$, where $\Gamma$ is the Euler gamma function. Strictly speaking, this quantity  measures entanglement between $A$ and its environment ($B$ + field). However, since the state of the field is quasi factorized at small coupling, it is also an approximate measure of entanglement between the two emitters $A$ and $B$. On one hand, purity~\eqref{purity} scales more slowly than the minimal value $(N+1)^{-1}$ in the sector, corresponding to maximally mixed reduced density matrices. On the other hand, this result is consistent with the minimal purity for states whose reduced density matrices are effectively approximated by the superposition of $O(N^{1/2})$ states. 

It is also possible to determine the entanglement properties of coherent and incoherent superpositions of the bound states $\ket{N}$. For example, one can consider the ``pseudothermal'' state 
\begin{equation}
\rho_{AB}^{\mathrm{th}} = (1-\mathrm{e}^{-\beta E}) \mathrm{e}^{-\beta E b_\phi^{\dagger}b_\phi} , 
\end{equation}
whose reduced density matrix is the thermal average of~(\ref{rhoAN}), yielding the purity
\begin{eqnarray}
\pi_{A}^{\mathrm{th}} & = & \sum_{\ell=0}^{\infty} \left( \sum_{N=\ell}^{\infty} (1-\mathrm{e}^{-\beta E}) \mathrm{e}^{-\beta E N} C_{\ell}^{(N)} (p_{\mathrm{at}}) \right)^2 \nonumber \\ & = & (1 - \mathrm{e}^{-\beta E} ) + O(N_{\mathrm{th}}\lambda^2),
\end{eqnarray}
with $N_{\mathrm{th}}$ the average excitation number. Another interesting case is the coherent state $\ket{\alpha} = \mathrm{e}^{\alpha b_\phi^{\dagger}-\alpha^* b_\phi} \ket{0}$, whose atom density matrix $\rho_{AB}^{\alpha}$ is dominated, in the small-coupling limit, by the projection on
\begin{equation}
\ket{\tilde{\alpha}} = \e^{-|\alpha|^2} \sum_{\ell=0}^{\infty} \frac{(-\alpha)^{\ell}}{\sqrt{2^{\ell}\ell!}} \sum_{m=0}^{\infty} \frac{(-\alpha)^{m}}{\sqrt{2^{m}m!}} \ket{ \ell_A, m_B },
\end{equation}
which yields factorized reduced density matrices.

Let us finally comment on the truncation of the oscillator Hilbert spaces. The bound state $\ket{N}$ contains up to the $N$th power of the oscillator creation operators $b_{A,B}^{\dagger}$. Therefore, truncating both oscillators at the $N$th level, namely formally adding the constraint $(b_{A,B}^{\dagger})^{N+1}=0$, has no effect on the existence and properties of state $\ket{N}$. On the other hand, the bosonic commutation relations satisfied by \textit{all} operators, which is reflected in the bosonic character of the $b_\phi$ operator, are fundamental in the derivation of the commutator equation~(\ref{ladder}), and in the subsequent reduction of the problem to the determination of $\phi_{A}$, $\phi_{B}$,  $\phi(k)$, and $E$. If the oscillators are truncated at $\bar{N}$, then the existence of a bound state in the sector $\mathcal{N}=N>\bar{N}$ does not follow automatically from the presence of a bound state in the lowest-excited $\mathcal{N}=1$ sector. Summarizing, in bound states, the excited levels of each oscillator must be in a sufficient number to absorb all the photons in the state. And, of course, considerations on the pseudothermal and coherent states are valid, only approximately, if the average occupation number is not close to $\bar{N}$.

\textit{ Conclusions and outlook.}---We have investigated the existence and properties of the stable states of a pair of $N$-level atoms in a waveguide, discussing the possibility to generate and preserve robust entanglement between the two atoms. 
As a case study, we have considered identical distinguishable emitters with uniform level spacing. The presence of (small or large) asymmetries in the excitation energy and in the coupling, or uneven levels, can be analyzed by following the approach suggested in~\cite{LNZ}. Among possible interesting applications, we mention the use of a single two-level emitter as a dynamical probe of the state of the field, along the guidelines discussed in~\cite{mirror1}. These ideas can be generalized, by using (entangled) atomic \emph{multilevel pairs}, paving the way to unprecedented possibilities and possibly super-resolution.

\begin{acknowledgments}
We thank F. Ciccarello, M. S. Kim, D. Pomarico and T. Tufarelli for early discussions.
PF and SP are partly supported by INFN through the project ``QUANTUM".
FVP is supported by INFN through the project ``PICS''.
PF is partially supported by the Italian National Group of Mathematical Physics (GNFM-INdAM).
This work is supported by the Top Global University Project from the Ministry of Education, Culture, Sports, Science and Technology (MEXT), Japan.
KY is supported by the Grant-in-Aid for Scientific Research (C) (No.~26400406) from the Japan Society for the Promotion of Science (JSPS) and by the Waseda University Grant for Special Research Projects (No.~2016K-215 and No.~2017K-236).
\end{acknowledgments}

\end{document}